\documentclass[a4paper,12pt]{article}

\usepackage{refmerge}
\usepackage[dvips]{graphicx}

\providecommand{\delopenpsi}[1][]{\ensuremath{{\scriptstyle\Delta}\psi_{#1}}}

\providecommand{\Ngamma}{\ensuremath{N_{\gamma}}}
\providecommand{\Egamma}[1][]{\ensuremath{E_{\gamma\,#1}}}

\providecommand{\Minv}[1][]{\ensuremath{M_{inv\,#1}}}

\providecommand{\etagppg}{\ensuremath{{\phi\rightarrow\eta\gamma},
\ {\eta\rightarrow\pi^+\pi^-\gamma}}}

\providecommand{\phietag}{\ensuremath{{\phi\rightarrow\eta\gamma}}}
\providecommand{\phipiog}{\ensuremath{{\phi\rightarrow\pi^0\gamma}}}

\providecommand{\etappg}{\ensuremath{{\eta\rightarrow\pi^+\pi^-\gamma}}}

\providecommand{\etappp}{\ensuremath{{\eta\rightarrow\pi^+\pi^-\pi^0}}}

\providecommand{\etaeeg}{\ensuremath{{\eta{\rightarrow}e^+e^-\gamma}}}

\providecommand{\phipioee}{\ensuremath{{\phi\rightarrow{\pi^0}e^+e^-}}}

\providecommand{\pioeeg}{\ensuremath{{\pi^0{\rightarrow}e^+e^-\gamma}}}

\providecommand{\phippp}{\ensuremath{{\phi\rightarrow\pi^+\pi^-\pi^0}}}

\providecommand{\pipipi}{\ensuremath{{\phi\rightarrow\pi^+\pi^-\pi^0}}}

\providecommand{\etagg}{\ensuremath{{\eta\rightarrow\gamma\gamma}}}
\providecommand{\piogg}{\ensuremath{{\pi^0\rightarrow\gamma\gamma}}}

\providecommand{\pioggg}{\ensuremath{{\phi\rightarrow\pi^0\gamma},
\ {\pi^0\rightarrow\gamma\gamma}}}

\providecommand{\eeg}{\ensuremath{e^+e^-\rightarrow e^+e^-\gamma}}

\providecommand{\eegg}{\ensuremath{e^+e^-\rightarrow e^+e^-\gamma\gamma}}

\providecommand{\MeV}{\mbox{MeV}}
\providecommand{\rad}{\mbox{radians}}

\providecommand{\cm}{\mbox{cm}}

\providecommand{\pb}{\mbox{pb}}

\title
{\Large
\bf \boldmath 
Observation of the conversion decay \phipioee\  at CMD-2
}

\author{
\parbox[c]{\textwidth}{\centering
R.R.~Akhmetshin\footnote{Budker Institute of Nuclear Physics,
Novosibirsk, 630090, Russia}, 
E.V.~Anashkin\footnotemark[1],
M.~Arpagaus\footnotemark[1],
V.M.~Aulchenko\footnotemark[1]\footnote{Novosibirsk State University, 
Novosibirsk, 630090, Russia}, 
V.Sh.~Banzarov\footnotemark[1],
L.M.~Barkov\footnotemark[1] \footnotemark[2],
N.S.~Bashtovoy\footnotemark[1], 
A.E.~Bondar\footnotemark[1] \footnotemark[2],
D.V.~Bondarev\footnotemark[1],
A.V.~Bragin\footnotemark[1], 
D.V.~Chernyak\footnotemark[1], 
S.I.~Eidelman\footnotemark[1] \footnotemark[2],
G.V.~Fedotovitch\footnotemark[1] \footnotemark[2], 
N.I.~Gabyshev\footnotemark[1], 
A.A.~Grebeniuk\footnotemark[1],
D.N.~Grigoriev\footnotemark[1], 
V.W.~Hughes\footnote{Yale University, New Haven, CT 06511, USA},
F.V.~Ignatov\footnotemark[1] \footnotemark[2],
P.M.~Ivanov\footnotemark[1], 
S.V.~Karpov\footnotemark[1],
V.F.~Kazanin\footnotemark[1] \footnotemark[2], 
B.I.~Khazin\footnotemark[1] \footnotemark[2],
I.A.~Koop\footnotemark[1],
P.P.~Krokovny\footnotemark[1] \footnotemark[2], 
L.M.~Kurdadze\footnotemark[1] \footnotemark[2], 
A.S.~Kuzmin\footnotemark[1] \footnotemark[2],
M.~Lechner\footnotemark[1],
I.B.~Logashenko\footnotemark[1], 
P.A.~Lukin\footnotemark[1],
K.Yu.~Mikhailov\footnotemark[1] \footnotemark[2],
I.N.~Nesterenko\footnotemark[1], 
V.S.~Okhapkin\footnotemark[1],
A.V.~Otboev\footnotemark[1],
E.A.~Perevedentsev\footnotemark[1] \footnotemark[2],
A.S.~Popov\footnotemark[1] \footnotemark[2],
T.A.~Purlatz\footnotemark[1] \footnotemark[2], 
N.I.~Root\footnotemark[1] \footnotemark[2],
A.A.~Ruban\footnotemark[1],
N.M.~Ryskulov\footnotemark[1],
A.G.~Shamov\footnotemark[1], 
Yu.M.~Shatunov\footnotemark[1],
B.A.~Shwartz\footnotemark[1] \footnotemark[2],
A.L.~Sibidanov\footnotemark[1] \footnotemark[2],
V.A.~Sidorov\footnotemark[1], 
A.N.~Skrinsky\footnotemark[1],
V.P.~Smakhtin\footnotemark[1],
I.G.~Snopkov\footnotemark[1], 
E.P.~Solodov\footnotemark[1] \footnotemark[2],
P.Yu.~Stepanov\footnotemark[1],
A.I.~Sukhanov\footnotemark[1], 
J.A.~Thompson\footnote{University of Pittsburgh, Pittsburgh, PA 15260, 
USA},
V.M.~Titov\footnotemark[1],
A.A.~Valishev\footnotemark[1], 
Yu.V.~Yudin\footnotemark[1],
S.G.~Zverev\footnotemark[1]
}
}

\begin{document}

\maketitle

\begin{abstract}
Using 15.1~$\mbox{pb}^{-1}$ of data collected by \mbox{CMD-2}
in the $\phi$-meson energy  range, the branching ratio of the conversion
decay \phipioee\  has been measured for the first time:
\begin{eqnarray}
 B(\phipioee) = (1.22 \pm 0.34 \pm 0.21) \cdot 10 ^{-5}. \nonumber
\end{eqnarray}
\end{abstract}

\section{Introduction}

Conversion decays of a vector meson $V$ into a pseudoscalar meson 
$P$ and a lepton pair $l^+l^-$ ($V \rightarrow P l^+l^-$, 
where $l=e,\mu$) are closely related to the corresponding radiative 
decays of $V$ into $P$ and a photon ($V\rightarrow P\gamma$). 

In conversion decays a squared invariant mass of a lepton pair
$\Minv^2(l^+l^-)=q^2$ or a squared mass of the virtual radiated
photon is not equal to zero as for usual radiative decays.
By studying  $\Minv(l^+l^-)$ spectra one can determine a
so called transition form factor $F_{P}(q^2_1,q^2_2)$ of 
pseudoscalar mesons $P$ as a function of $q^2_{i}$.

Various theoretical models, from standard vector meson dominance 
(VMD) to calculations on the lattice, predict the transition form factor
and the resulting branching ratio of the decay
\cite{landsberg,bramon,lattice,faes}.
 
Precise values of the branching ratios of conversion decays are
also important while studying the yield of direct
lepton pairs in heavy ion collisions. As it was noted long ago, 
an observation of the anomalously large yield could 
indicate the existence of the quark-gluon plasma  \cite{shuryak}.
In general, both the yield of dileptons and their  mass spectra are
of interest while studying the change of the meson properties
in medium \cite{chin}. 
Recent experiments studying dileptons in heavy ion collisions at
CERN reported on an excess of the number of $e^+e^-$ \cite{ceres_taps} 
and $\mu^+\mu^-$ pairs \cite{helios} over the expectations from usual 
hadron decays ($\pi^0 \to e^+e^-\gamma$, $\eta \to e^+e^-\gamma$, 
$\eta^{\prime}\to e^+e^-\gamma$, $\omega\to\pi^0e^+e^-$, 
$\rho/\omega\to e^+e^-$, $\phi\to e^+e^-$
as well as similar decays into muon pairs). Its explanation requires 
the accurate calculation of the production rate of pseudoscalar and 
vector mesons as well as good knowledge of both the invariant mass 
spectra and the branching ratios of their decays into lepton pairs.

The experimental information on such decays is rather scarce \cite{pdg}.
While for the $\omega$ meson both possible decays into $\pi^0$ were 
observed, $\omega \to \pi^0 \mu^+\mu^-$ \cite{lande} and 
$\omega \to \pi^0 e^+e^-$ \cite{nd1},
only few events of the decay $\phi \to \eta e^+e^-$ were 
detected \cite{nd2}.
  
A large data sample of the $\phi$ mesons collected by two detectors
at \mbox{VEPP-2M} allowed reliable detection of the decay mode 
$\phi \to \eta e^+e^-$ by both CMD-2 \cite{e1} and SND \cite{e2}
groups. CMD-2 has also reported on the first observation of the 
$\phi \to \pi^0 e^+e^-$ decay \cite{e1}.
  
This work is devoted to the determination of the branching ratio for the
conversion decay \phipioee\  using the complete data sample available at
CMD-2.


\section{Experiment}

The general purpose detector \mbox{CMD-2} installed at the 
$e^+e^-$ collider \mbox{VEPP-2M} \cite{vepp} has been described in 
detail elsewhere \cite{cmddec}. 

It consists of a cylindrical drift chamber (DC) and double-layer
multi-wire proportional Z-chamber, both also used for the trigger, and
both inside a thin (0.38 $X_0$) superconducting solenoid with a field
of 1T. 
The momentum resolution of the DC is equal to
$\sigma_p / p = \sqrt{90 \cdot (p(GeV))^2+7}\%$ . 
The accuracy in the measurement of polar and azimuthal angles is
$\sigma_{\theta}=1.5 \cdot 10^{-2}$ and $\sigma_{\phi} = 7 \cdot
10^{-3}$ radians respectively.

The barrel calorimeter with a thickness of 8.1$X_0$ is placed outside
the solenoid and consists of 892 CsI crystals.
The energy resolution for photons is about 9\% in the energy range
from 50 to 600 MeV.  
The angular resolution is of the order of 0.02 radians.

The end-cap calorimeter placed inside the solenoid consists of 680 BGO
crystals. 
The thickness of the calorimeter for normally incident particles is
equal to 13.4$X_0$.  
The energy and angular resolution varies from 8\% to 4\% and from 0.03
to 0.02 radians respectively for the photon energy in the range 
100 to 700 MeV.
Both barrel and end-cap calorimeters cover a solid angle of
0.92$\times4\pi$ steradians. 

The experiment was performed in the $\phi$ meson energy range 
(985-1060 MeV). The integrated luminosity collected
during the runs of 1993 (PHI93), 1996 (PHI96) and 1998 (PHI98)
was 1.5, 2.1 and 11.5~$\pb^{-1}$ respectively, so that our
analysis is based on the data sample corresponding to
15.1~$\pb^{-1}$.

\section{Data analysis}
\subsection{General approach}

A search for events of the decay \phipioee\  used 
the final state with two charged particles and two photons from
the decay $\pi^0 \to \gamma \gamma$. For the process under study 
the invariant mass of the two photons $\Minv(\gamma\gamma)$ 
should  be equal to
the $\pi^0$-mass within the resolution.
Event selection was performed using kinematical reconstruction taking
into account energy-momentum conservation.
The number of detected events is given by:
\begin{equation}
\label{eq:n_pioeegg}
N_{\phipioee} = N_{\phi} \cdot B(\phipioee) \cdot B(\piogg) \cdot
\varepsilon_{\phipioee}
\end{equation}
where $N_{\phi}$ is the total number of the produced $\phi$ mesons
and $\varepsilon_{\phipioee}$ is the corresponding detection
efficiency.

One of the most important backgrounds to our process comes from events 
of the decay \pioggg\  followed by the $\gamma$-quantum conversion 
at the wall of the beam pipe.
Since the DC space resolution is not sufficient to separate events
with the conversion at the beam pipe from those due to conversion decays
at the interaction point, the contribution of 
this background was calculated from the simulation:
\begin{equation}
\label{eq:n_pioeegg_pioggg}
N_{\phipiog} = N_{\phi} \cdot B(\phipiog) \cdot B(\piogg)
\cdot \varepsilon_{\phipiog},
\end{equation}
where $\varepsilon_{\phipiog}$ is the detection efficiency of the
\phipiog\  decay with the $\gamma$ conversion in the material.

Another significant source of background is the decay mode \phippp.
In this case the kinematical
reconstruction assuming that charged tracks are electrons, 
gives $\Minv(\gamma\gamma)$ about 150~MeV, i.e.
only slightly different from the $\pi^0$-mass, and
is not sufficient to separate this process from 
(\ref{eq:n_pioeegg}). 
Therefore, to suppress this background, the procedure of $e/\pi$ 
separation by the energy loss of charged particles in the calorimeter 
was developed (see the next subsection and \cite{rho4pi}) allowing
determination of the expected number $N_{\phi \to \pi^+\pi^-\pi^0}$ 
of such events.

Other sources of background are the quantum electrodynamical (QED)
process \eegg, and the decay mode \phietag\  followed by  
the Dalitz decay \etaeeg\  or the conversion of $\gamma$
from the \etagg\  decay.
All these background processes can be separated from the
decay under study by their wide and smooth distribution of
$\Minv(\gamma\gamma)$. 

The total number of observed events is a sum of the three 
contributions above:
\begin{equation}
\label{eq:n_pioeegg_exp}
N_{\phipioee}^{exp} = N_{\phipioee} + N_{\phipiog} +
N_{\pipipi}.
\end{equation}


The following selection criteria were used for the decay \phipioee:
\begin{itemize}
\item the impact parameter of the tracks $d<0.25~\cm$ to
reject events with the $K^0_{S,L}$ decay in DC;
\item the angle between two tracks $\delopenpsi < 0.3~\rad$ to
suppress the process \phippp ~(for conversion decays the angle
between $e^+$ and $e^-$ is small); 
\item the number of photons $\Ngamma^{KREC} = 2$;
\item the photon energy $\Egamma^{max} < 460~\MeV$ to suppress
the process \eegg;
\item $\Egamma^{min} > 50~\MeV$ for calorimeter noise suppression;
\item $|\Egamma^{max}-362|>15~\MeV$ for suppression of the conversion
decay \etaeeg.
\end{itemize}

To determine the total number of the produced $\phi$ mesons,
events of the process \etagppg\  have been used. 
Since this process also contains two charged particles and two photons in
the final state, normalization to it allows to cancel some
possible  systematic errors.
The number of events of this process is given by the following formula: 
\begin{equation}
\label{eq:n_phi_etagppg}
N_{\etappg} = N_{\phi} \cdot B(\phietag) \cdot B(\etappg) \cdot
\varepsilon_{\etappg}.
\end{equation}

For the normalization process the selection criteria are:
\begin{itemize}
\item $\delopenpsi < 2.5$  to suppress 
\mbox{$K^0_S\to\pi^+\pi^-$} events;
\item $\Ngamma^{KREC} = 2$;
\item $\Minv(\gamma\gamma) > 250$ MeV for
suppression of \phippp\  events;
\item $d<0.25~\cm$ to reject events with the $K^0_{S,L}$ decay in DC.
\end{itemize}

\begin{figure}[ht]
\centering
\includegraphics[width=0.47\textwidth]{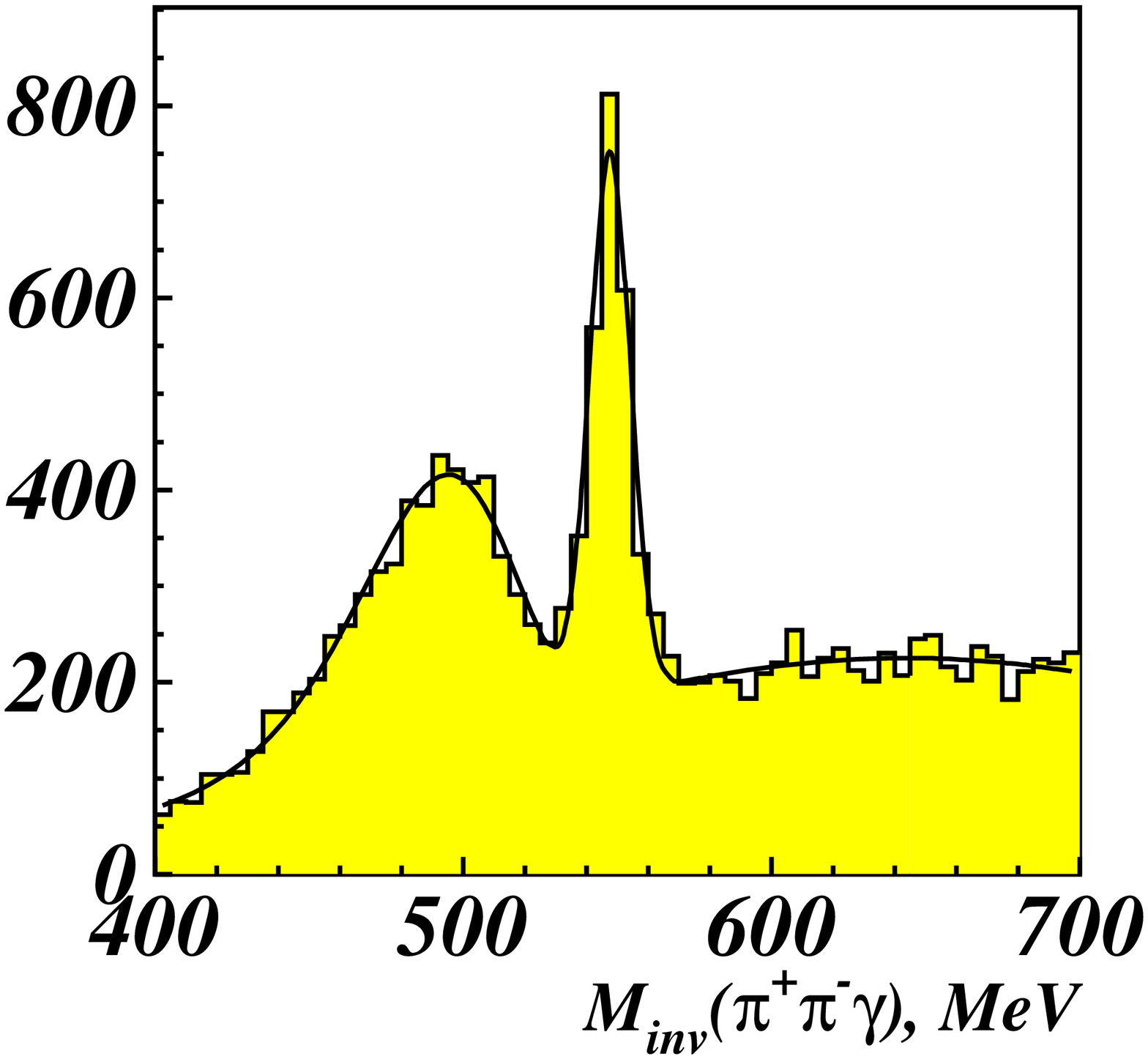}
\caption{Invariant mass $\Minv(\pi^+\pi^-\gamma)$ for \phietag, \etappg} 
\label{fig:etagppg_phi98}
\end{figure}

Figure~\ref{fig:etagppg_phi98} shows the distribution of the invariant 
mass $\Minv(\pi^+\pi^-\gamma)$ for thus selected experimental events. 
A clear signal from the process \phietag, \etappg\  is observed at the
$\eta$-meson mass. It can be fit with a  Gaussian whose
variance  is taken from simulation.
The background mainly comes from events of the process \phietag, \etappp\
with a lost photon (the wide signal in the 500~\MeV\  region can be
fit with the logarithmic Gaussian \cite{handbook}), the decay \phippp\  
and the QED process
\mbox{\eegg} (another wide background fit with the Gaussian). 

Eliminating $N_{\phi}$ from the relations (1), (2) and (4), the following
expression for the branching ratio of the decay \phipioee\
can be obtained from (3):
\begin{equation}
\label{eq:b_pioeegg_etagppg}
\begin{array}{c}
B(\phipioee) = \\
B(\phietag) \cdot
\frac{\displaystyle\mathstrut B(\etappg)}
{\displaystyle\mathstrut B(\piogg)} \cdot
\frac{\displaystyle\mathstrut \varepsilon_{\etappg}}
{\displaystyle\mathstrut \varepsilon_{\phipioee}} \cdot
\frac{\displaystyle\mathstrut N_{\phipioee}^{exp}~-~N_{\phippp}}
{\displaystyle\mathstrut N_{\etappg}} \\
- B(\phipiog) \cdot
\frac{\displaystyle\mathstrut\varepsilon_{\phipiog}}
{\displaystyle\mathstrut\varepsilon_{\phipioee}}
\end{array}
\end{equation}
Here $N_{\pipipi}$ is the expected number of the events of the
decay $\phi \to \pi^+\pi^-\pi^0$ obtained from the procedure of $e/\pi$ 
separation.



\subsection{\boldmath $e/\pi$-separation}

Since simulation can not completely reproduce interactions of charged
particles with the calorimeter, the procedure of $e/\pi$-separation is
based on the experimental data.
Energy losses of $e^{\pm}$ and $\pi^{\pm}$  were studied
using ``clean'' samples of events of
the processes \eeg\  and \phippp, \piogg\  respectively.
The QED process \eeg\  with small angles between
the tracks was chosen as the closest to \phipioee.

\begin{figure}[ht]
\includegraphics[width=0.47\textwidth]{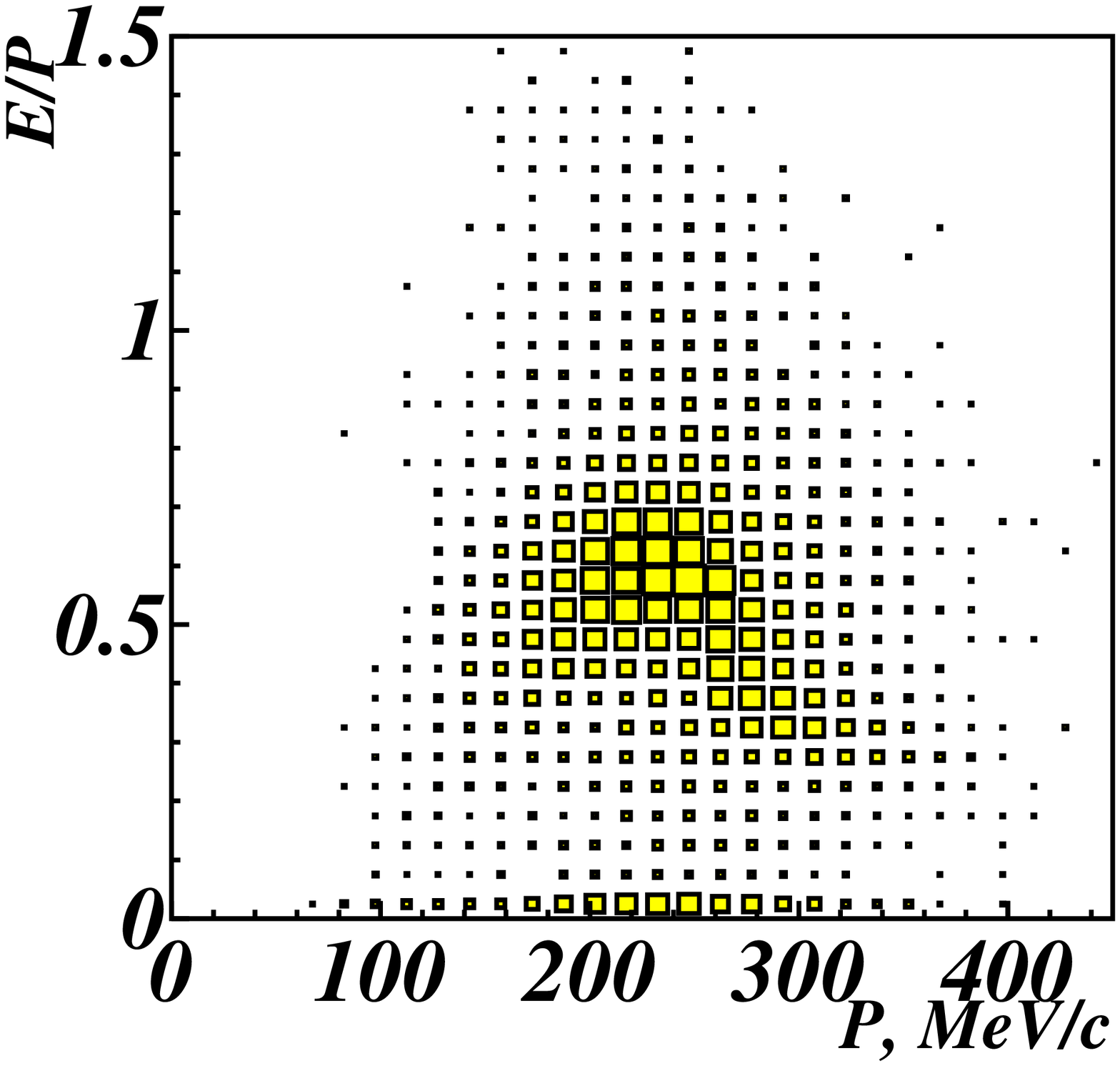}
\hfill
\includegraphics[width=0.47\textwidth]{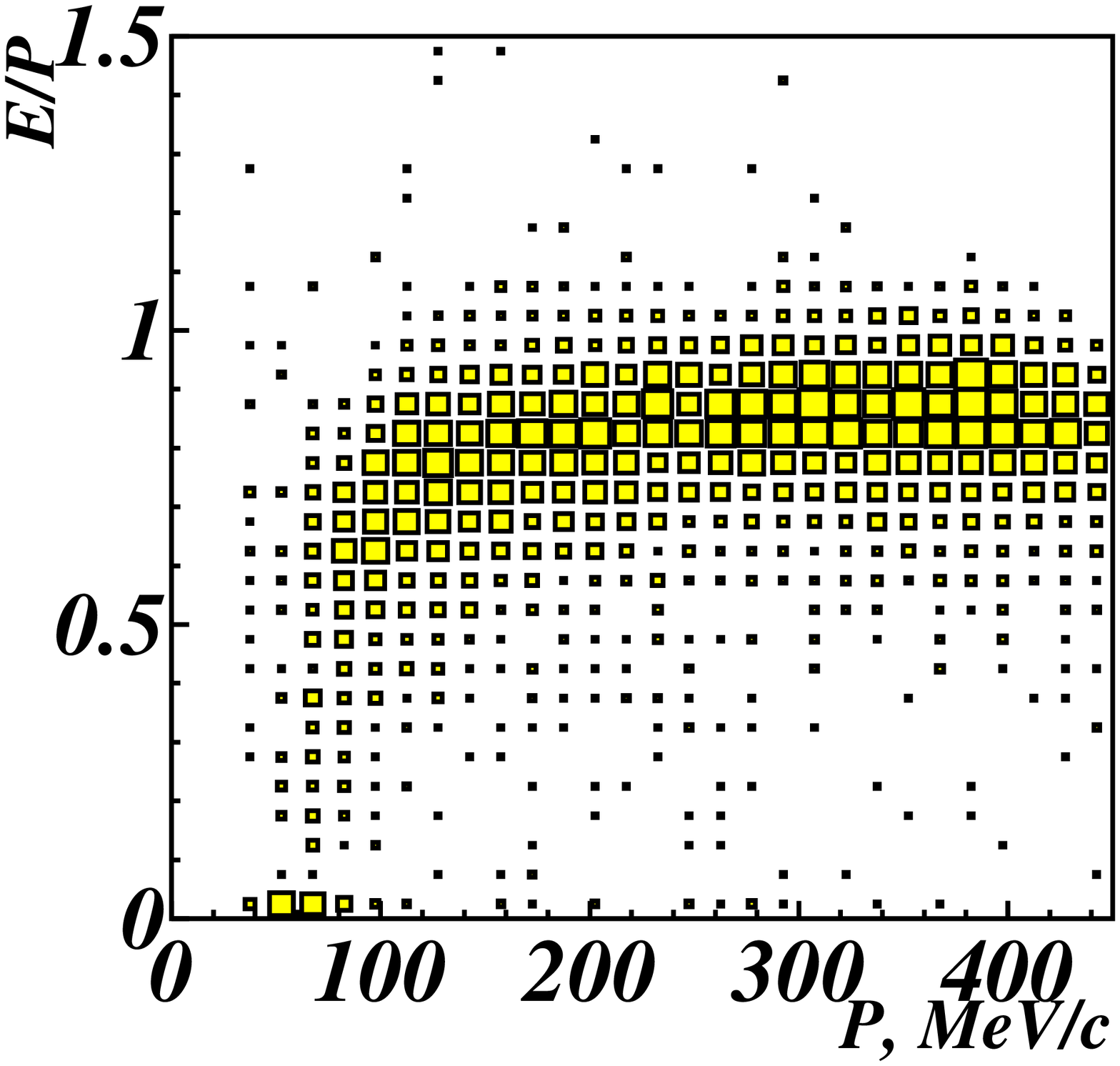}
\\
\parbox[t]{0.47\textwidth}{\vspace{-1.5cm}\center $a)$ $\pi^+$}
\hfill
\parbox[t]{0.47\textwidth}{\vspace{-1.5cm}\center $b)$ $e^+$}
\vspace{-1.0cm}
\caption{Distribution of $E/P$ vs $P$: 
$a)$ $\pi^+$ for \phippp; $b)$ $e^+$ for \eeg} 
\label{fig:ecltpt_vs_pt}
\end{figure}

From the distributions of $E/P$ (the energy loss
of a charged particle in the calorimeter divided by its momentum in DC)
versus $P$ shown in Fig.~\ref{fig:ecltpt_vs_pt}~$a)$ for pions from
the \phippp\  decay and in Fig.~\ref{fig:ecltpt_vs_pt}~$b)$ for electrons
from the \eeg\,  it can be seen that $e/\pi$ can be separated well.

The whole region of momenta from 0 to 500~MeV was divided to bins 50~MeV
wide and for each bin the $E/P$ distributions were plotted.
These spectra were fit by: for $e^{\pm}$ -- the logarithmic Gaussian
function, for $\pi^{\pm}$ -- a sum of two standard Gaussian functions.
A probability to penetrate the detector without any energy loss and a 
probability to lose some energy in the calorimeter were defined
as a ratio of the number of events with $E=0$ and with $E>0$ to 
the total number of events respectively.

Then the probability density functions for each sort of the particles
$f_{e^{\pm},\pi^{\pm}}$ depend on the momentum in DC and energy loss in
the calorimeter. 

The probability for one track to be $\pi^+$ or $\pi^-$:
\begin{equation}
\label{eq:W_pi}
W_{\pi^+}=\frac{f_{\pi^+}}{f_{\pi^+}+f_{e^+}}, ~
W_{\pi^-}=\frac{f_{\pi^-}}{f_{\pi^-}+f_{e^-}}
\end{equation}

The probability for one track to be $e^+$ or $e^-$:
\begin{equation}
\label{eq:W_e}
W_{e^+}=\frac{f_{e^+}}{f_{\pi^+}+f_{e^+}}, ~
W_{e^-}=\frac{f_{e^-}}{f_{\pi^-}+f_{e^-}}
\end{equation}

Similarly, one can construct the probability for both tracks in the
event to be $\pi^+\pi^-$ or $e^+e^-$:
\begin{equation}
\label{eq:W_pp_ee}
W_{\pi^+\pi^-}=\frac{W_{\pi^+} \cdot W_{\pi^-}}
{W_{\pi^+} \cdot W_{\pi^-}+W_{e^+} \cdot W_{e^-}}, ~
W_{e^+e^-}=\frac{W_{e^+} \cdot W_{e^-}}
{W_{\pi^+} \cdot W_{\pi^-}+W_{e^+} \cdot W_{e^-}}
\end{equation}


Finally, we can determine the detection efficiencies
$\varepsilon^{e/\pi}_{e^+e^-}$ and $\varepsilon^{e/\pi}_{\pi^+\pi^-}$ 
which respectively give the probability to detect an $e^+e^-$ and 
$\pi^+\pi^-$ pair as an $e^+e^-$ one. They are calculated as the 
ratio of the number of events with $W_{e^+e^-}>W_{0}$ to the total 
number of sample events of the processes \eeg\  and \pipipi\ 
respectively. Here $W_{0}$ is some boundary value chosen from the 
analysis of the $W_{e^+e^-}$ distribution.

\subsection{\boldmath Selection of \phipioee\  events}

The main selection criteria for the process under study were listed above.

\begin{figure}[ht]
\vspace{-0.25cm}
\centering
\includegraphics[width=0.47\textwidth]{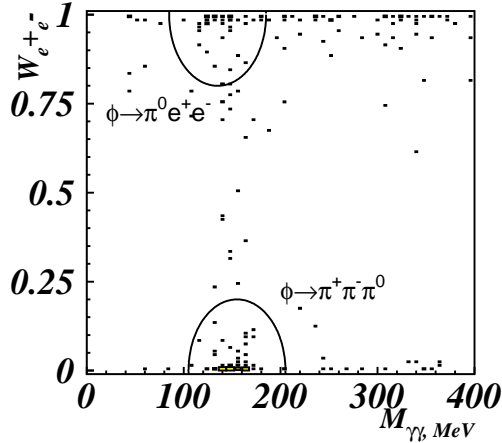}
\caption{Scatter plot of $W_{e^+e^-}$ versus $\Minv(\gamma\gamma)$}
\label{fig:ecltpt_vs_pioeegg_phi98}
\vspace{-0.25cm}
\end{figure}

Figure~\ref{fig:ecltpt_vs_pioeegg_phi98} shows 
the scatter plot of the probability  $W_{e^+e^-}$
for both tracks to be electrons
versus the invariant mass of the photons
$\Minv(\gamma\gamma)$ for the PHI98 data after these cuts.
Events with an $e^+e^-$-pair have $W_{e^+e^-}\sim1$ while for those 
with pions $W_{e^+e^-}\sim0$. 
Events in the region $W_{e^+e^-}\sim0$ and
$\Minv(\gamma\gamma)\sim150$~MeV come from the decay  \phippp.
To suppress it, the additional cut based on the $e/\pi$-separation 
was applied:
\begin{itemize}
\item \mbox{$W_{e^+e^-}>0.5$}, the probability that an event 
has an $e^+e^-$-pair.
\end{itemize}

\begin{figure}[ht]
\centering
\includegraphics[width=0.47\textwidth]{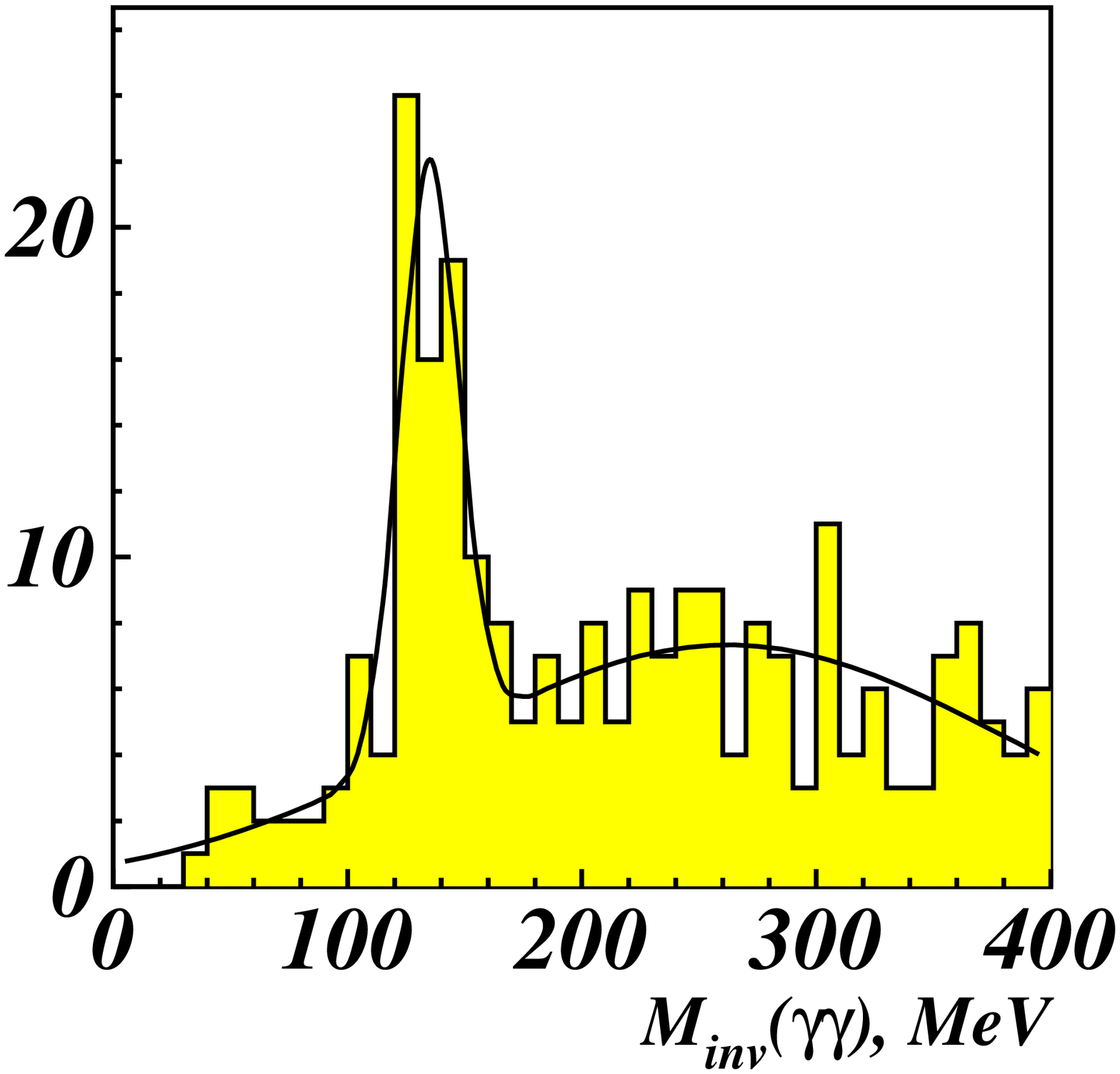}
\caption{The invariant mass $\Minv(\gamma\gamma)$ for \phipioee, \piogg} 
\label{fig:pioeegg_phi98}
\end{figure}

Figure~\ref{fig:pioeegg_phi98} shows the distribution of the
invariant mass $\Minv(\gamma\gamma)$ for the data after this selection.
A clear signal is observed at the $\pi^0$-meson mass which
can be fit with a Gaussian (its variance is taken from simulation).
The wide background can be also fit with a Gaussian.

If instead the cut $W_{e^+e^-} > 0.5$ one applies the complementary cut
\begin{itemize}
\item \mbox{$W_{e^+e^-}<0.5$}, to enhance the probability that the 
event has a $\pi^+\pi^-$ pair,
\end{itemize}
the expected number $N_{\phi \to \pi^+\pi^-\pi^0}$ can be determined. 

Detection efficiencies were determined from the Monte Carlo simulation
(MC) \cite{cmd2sim} 
taking into account the efficiencies of $e/\pi$ separation 
$\varepsilon_{e^+e^-}^{e/\pi}, \varepsilon_{\pi^+\pi^-}^{e/\pi}$ and
a correction $\varepsilon_{\Delta\Psi}$ 
for the reconstruction of events with a small opening angle,
for brevity referred to as the small angle correction. 
The reconstruction
of such events can not be completely reproduced by the simulation
and was therefore specially studied using experimental events of the
process \phippp, \pioeeg\  which also have a small angle between an
electron and positron.
$$
\varepsilon_{\phipioee} = \varepsilon_{\phipioee}^{MC} \cdot
\varepsilon_{\Delta\Psi} \cdot \varepsilon_{e^+e^-}^{e/\pi},~  
\varepsilon_{\phipiog} = \varepsilon_{\phipiog}^{MC} \cdot
\varepsilon_{\Delta\Psi} \cdot \varepsilon_{e^+e^-}^{e/\pi},
$$
$$
\varepsilon_{\pipipi} = \varepsilon_{\pipipi}^{MC} \cdot
\varepsilon_{\Delta\Psi} \cdot \varepsilon_{\pi^+\pi^-}^{e/\pi}.
$$

\section{Results}
\begin{figure}[ht]
\centering
\includegraphics[width=0.47\textwidth]{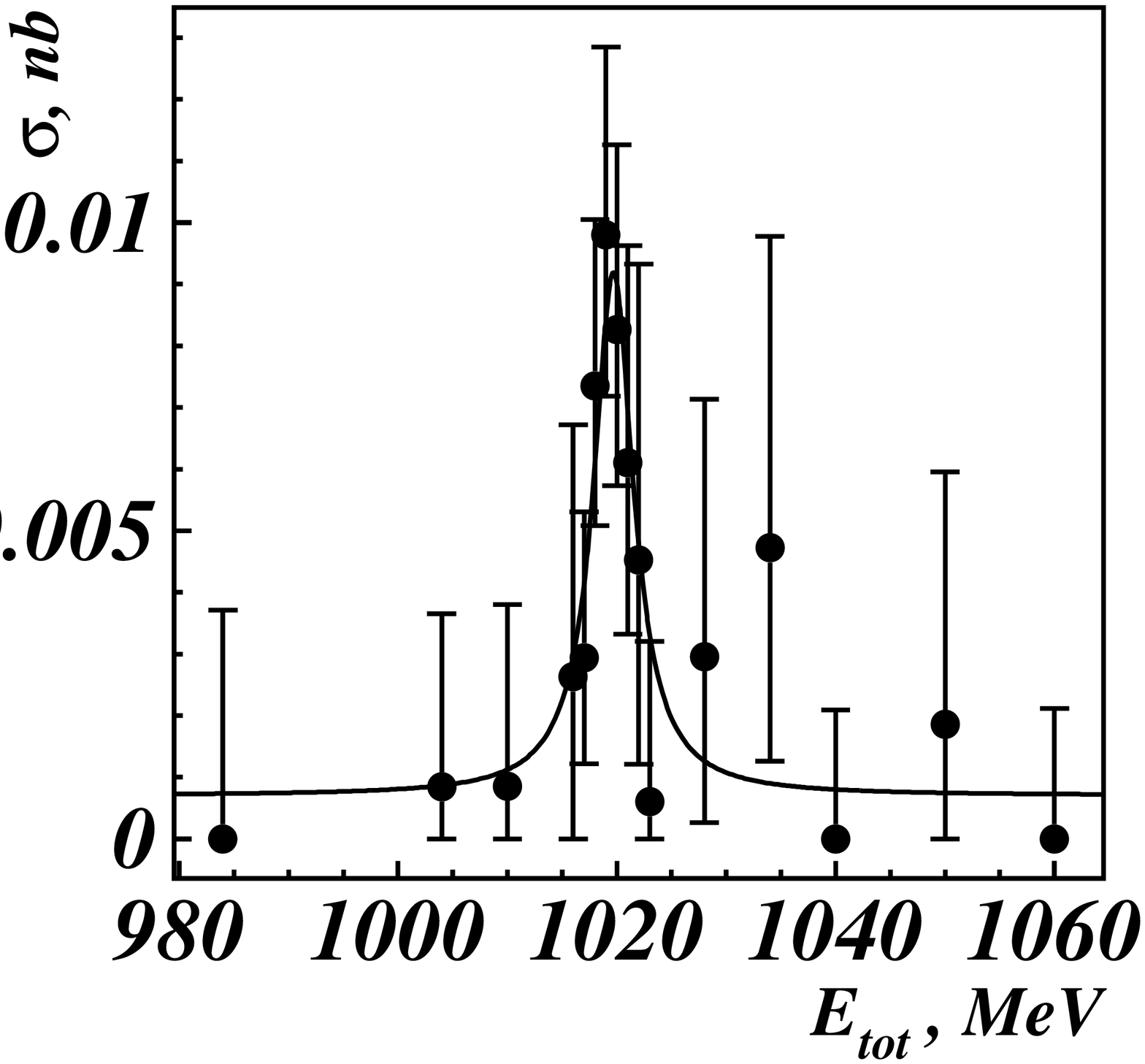}
\caption{The visible cross section of \phipioee, \piogg\
versus the total energy} 
\label{fig:phi_res_pioeegg_phi98}
\end{figure}

Figure~\ref{fig:phi_res_pioeegg_phi98} shows the 
energy dependence of the visible cross section of
the process  \phipioee, \piogg\  for the run PHI98. 
It is compatible with that expected for the $\phi$ 
meson, but a small non-resonant background can not be excluded.
In fact, one should expect such a background coming from the
''tail'' of the $\omega \to \pi^0e^+e^-$ decay. To determine
the number of events coming from the $\phi$ meson decay, a fit
of the visible cross section was performed which included
a Breit-Wigner signal of the $\phi$ meson and a possible
non-resonant background. The fit with the $\phi$ meson
mass and width fixed at their world average values gave
the following 
ratio of the non-resonant background cross section
to that at the peak
$$
\frac{\sigma_{\omega}}{\sigma_{\phi}}~=~7.5^{+15.4}_{~-7.5}\%
$$
compatible with the VMD estimate.
From this value as well as from the integrated luminosity one
can determine the corrected number of events for the process under 
study:
$$
N_{\phi \to \pi^0e^+e^-}^{exp}~=~47.6 \pm 11.1.
$$
Similar analysis was performed for the data of PHI93 and PHI96 runs.


Table~\ref{tab:pioeegg_pid_dpst_phi_all} presents the final results
of the data processing: the number of selected
events, the expected number of events of \phipioee,
\phipiog\  and \phippp\  determined by the described procedure, 
the detection efficiencies and branching ratios
of the \phipioee\  decay in various runs.
Also presented in the Table are the data for the normalization process.
The differences of the efficiencies from run to run are 
caused by varying conditions of data taking.
The total detected number of the  \phipioee\ candidates  is 
$67.9 \pm 12.3$ with an expected background of $22.3\pm1.4$ events. 
The error in the estimated number of background events 
is determined by the procedure described above.

\begin{table}[ht]
\caption{Branching ratio of \phipioee\  decay}
\label{tab:pioeegg_pid_dpst_phi_all}
\begin{center}
\begin{tabular}{lcccc}
\hline
\hline
Run & PHI93 & PHI96 & PHI98 & Total \\
\hline
\hline
$N_{\phipioee}^{exp}$ & 
\hspace{-5mm} \begin{tabular}{c} $~6.3\pm~2.8$ \\ $\pm0.1$
\end{tabular} \hspace{-5mm} &  
\hspace{-5mm} \begin{tabular}{c} $14.1\pm~4.4$ \\ $\pm0.5$
\end{tabular} \hspace{-5mm} &  
\hspace{-5mm} \begin{tabular}{c} $47.6\pm~11.1$ \\ $\pm3.6$
\end{tabular} \hspace{-5mm} & 
\hspace{-5mm} \begin{tabular}{c} $67.9\pm12.3$ \\ $\pm5.2$
\end{tabular} \hspace{-5mm} \\ 
\hline
$\varepsilon_{\pi^0e^+e^-}^{MC}, \%$ & 
\hspace{-5mm} \begin{tabular}{c} $19.11\pm0.31$ \\ $\pm2.18$
\end{tabular} \hspace{-5mm} & 
\hspace{-5mm} \begin{tabular}{c} $26.74\pm0.37$ \\ $\pm2.51$
\end{tabular} \hspace{-5mm} & 
\hspace{-5mm} \begin{tabular}{c} $23.10\pm0.34$ \\ $\pm1.82$
\end{tabular} \hspace{-5mm} \\
$\varepsilon_{\pi^0\gamma}^{MC}, 10^{-4}$ & 
\hspace{-5mm} \begin{tabular}{c} $4.98\pm0.22$ \\ $\pm0.40$
\end{tabular} \hspace{-5mm} & 
\hspace{-5mm} \begin{tabular}{c} $7.30\pm0.27$ \\ $\pm0.59$
\end{tabular} \hspace{-5mm} & 
\hspace{-5mm} \begin{tabular}{c} $5.44\pm0.23$ \\ $\pm0.44$
\end{tabular} \hspace{-5mm} & \\ 
\hline
$\varepsilon_{\Delta\Psi}, \%$ & 
\hspace{-5mm} \begin{tabular}{c} $99.5\pm20.1$ \\ $\pm7.1$
\end{tabular} \hspace{-5mm} & 
\hspace{-5mm} \begin{tabular}{c} $91.0\pm10.4$ \\ $\pm5.6$
\end{tabular} \hspace{-5mm} & 
\hspace{-5mm} \begin{tabular}{c} $96.4\pm~2.8$ \\ $\pm4.7$
\end{tabular} \hspace{-5mm} & \\ 
\hline
$\varepsilon_{e^+e^-}^{e/\pi}, \%$ & 
$92.4\pm8.3$ & $93.3\pm5.4$ & $89.9\pm1.3$ & \\
$\varepsilon_{\pi^+\pi^-}^{e/\pi}, \%$ & 
$9.1\pm2.9$ & $7.8\pm1.0$ & $7.0\pm0.4$ & \\
\hline
\hspace{-5mm} \begin{tabular}{c} Expected \\ $N_{\phipiog}$
\end{tabular} \hspace{-5mm} &  
\hspace{-5mm} \begin{tabular}{c} $~0.9\pm0.3$ \\ $\pm0.1$
\end{tabular} \hspace{-5mm} &  
\hspace{-5mm} \begin{tabular}{c} $~2.2\pm0.3$ \\ $\pm0.3$
\end{tabular} \hspace{-5mm} &  
\hspace{-5mm} \begin{tabular}{c} $~8.5\pm0.6$ \\ $\pm1.1$
\end{tabular} \hspace{-5mm} &  
\hspace{-5mm} \begin{tabular}{c} $11.7\pm0.7$ \\ $\pm1.5$
\end{tabular} \hspace{-5mm} \\ 
\hspace{-5mm} \begin{tabular}{c} Expected \\ $N_{\pi^+\pi^-\pi^0}$
\end{tabular} \hspace{-5mm} &  
\hspace{-5mm} \begin{tabular}{c} $~1.4\pm0.6$ \\ $\pm0.1$
\end{tabular} \hspace{-5mm} &  
\hspace{-5mm} \begin{tabular}{c} $~1.7\pm0.5$ \\ $\pm0.1$
\end{tabular} \hspace{-5mm} &  
\hspace{-5mm} \begin{tabular}{c} $~7.4\pm1.0$ \\ $\pm0.1$
\end{tabular} \hspace{-5mm} &  
\hspace{-5mm} \begin{tabular}{c} $10.6\pm1.3$ \\ $\pm0.1$
\end{tabular} \hspace{-5mm} \\ 
\hspace{-5mm} \begin{tabular}{c} Total expected \\ background 
\end{tabular} \hspace{-5mm} & 
\hspace{-5mm} \begin{tabular}{c} $~2.3\pm0.7$ 
\end{tabular} \hspace{-5mm} &  
\hspace{-5mm} \begin{tabular}{c} $~4.0\pm0.6$ 
\end{tabular} \hspace{-5mm} &  
\hspace{-5mm} \begin{tabular}{c} $16.0\pm1.1$ 
\end{tabular} \hspace{-5mm} &  
\hspace{-5mm} \begin{tabular}{c} $22.3\pm1.4$ 
\end{tabular} \hspace{-5mm} \\ 
\hspace{-5mm} \begin{tabular}{c} Expected \\ $N_{\phipioee}$
\end{tabular} \hspace{-5mm} &  
\hspace{-5mm} \begin{tabular}{c} $~3.9\pm2.9$ 
\end{tabular} \hspace{-5mm} &  
\hspace{-5mm} \begin{tabular}{c} $10.1\pm4.4$ 
\end{tabular} \hspace{-5mm} &  
\hspace{-5mm} \begin{tabular}{c} $31.6\pm11.2$ 
\end{tabular} \hspace{-5mm} &  
\hspace{-5mm} \begin{tabular}{c} $45.6\pm12.4$ 
\end{tabular} \hspace{-5mm} \\ 
\hline
$\varepsilon_{\etappg}, \%$ & 
$12.34\pm0.25$ & $20.47\pm0.32$ & $20.75\pm0.32$ & \\
\hspace{-5mm} \begin{tabular}{c}   $N_{\etappg}$ 
\end{tabular} \hspace{-5mm} &   
\hspace{-5mm} \begin{tabular}{c} $~126\pm17$ \\ $\pm~6$ 
\end{tabular} \hspace{-5mm} &  
\hspace{-5mm} \begin{tabular}{c} $~362\pm25$ \\ $\pm~9$ 
\end{tabular} \hspace{-5mm} & 
\hspace{-5mm} \begin{tabular}{c} $1858\pm58$ \\ $\pm38$ 
\end{tabular} \hspace{-5mm} & 
\hspace{-5mm} \begin{tabular}{c} $2346\pm65$ \\ $\pm53$ 
\end{tabular} \hspace{-5mm} \\
\hline
\hline
\hspace{-5mm} \begin{tabular}{c}
$B(\phipioee),$ \\ $10^{-5}$ 
\end{tabular} \hspace{-5mm} & 
\hspace{-5mm} \begin{tabular}{c} 
$1.36\pm1.13$ \\ $\pm0.22$
\end{tabular} \hspace{-5mm} &   
\hspace{-5mm} \begin{tabular}{c}
$1.57\pm0.75$ \\ $\pm0.23$
\end{tabular} \hspace{-5mm} &   
\hspace{-5mm} \begin{tabular}{c}
$1.10\pm0.40$ \\ $\pm0.19$
\end{tabular} \hspace{-5mm} &   
\hspace{-5mm} \begin{tabular}{c}
$1.22\pm0.34$ \\ $\pm0.21$
\end{tabular} \hspace{-5mm} \\
\hline
\hline
\end{tabular}
\end{center}
\end{table}

As already noted, since the number of events of the process under 
study is normalized to that of the process 
$\phi \to \eta \gamma$, \etappg\  with a similar
final state, some of the systematic uncertainties cancel (trigger, 
detection efficiencies etc.). 

The main sources of the remaining systematic errors are listed below, 
their magnitude given for the most statistically significant run
PHI98:
\begin{itemize}
\item
A limited sample of simulated events used to determine detection
efficiencies - 2.7\%
\item
Statistical errors of the parameters in the small angle
correction - 3.9\%
\item
Parameters of the $e/\pi$ separation procedure - 2.5\%
\item
The shape of the invariant mass
distributions used in the fits - 11.9\% 
\item 
Inaccurate knowledge of the thickness of
material in front of the DC (the beryllium beam pipe and 
the DC inner wall made of aluminized mylar) - 2.2\% 
\item
The shape of the distributions used in the fits to determine the 
parameters of the small angle correction - 6.8\%
\item
Errors of the world average values for the branching ratios
of intermediate decays from \cite{pdg} - 4.9\%
\item
Dependence on the transition form factor model. The form factor in the
generalized VMD ($\rho + \rho'$ mesons) was compared to that in 
the simple VMD (single $\rho$ meson only). The contribution to
the branching ratio is 9.5\%. 
\end{itemize}

Three first sources of the error are of statistical nature, their
values were obtained for each run separately, so that they 
are uncorrelated. Their total contribution to the error is
5.4\% and can be quadratically added to the statistical error. 
All others are obviously correlated from 
run to run and their total magnitude is 17.5\%.  

From the last line of Table 1 it is clear that the values of the
branching ratio obtained in different runs are consistent within
the errors. To improve the accuracy, we can average them. 
The resulting value of the branching ratio for the \phipioee\ 
decay is  
\mbox{$(1.22 \pm 0.34 \pm 0.21) \cdot 10^{-5}$}
where the first error is statistical (including the uncorrelated
systematic errors discussed above) and the second one is a
systematic correlated error common for all runs.

\section{Conclusion}
For the first time the branching ratio of the conversion decay 
$\phi \to \pi^0 e^+e^-$ has been
determined by the CMD-2 detector at VEPP-2M:
\begin{equation}
B(\phi \to \pi^0 e^+e^-) = (1.22\pm0.34\pm0.21)\cdot10^{-5}.
\label{res:b_phipioee_cmd} 
\end{equation}
This measurement is based on 68 selected candidates for the events of
the process \phipioee\  with an expected background of 22 events.
The obtained value agrees with the theoretical predictions
\mbox{$(1.3-1.6)\cdot10^{-5}$} \cite{faes,eidelman,prediction} and with
the experimental upper limit \mbox{$1.2\cdot10^{-4}$} at 90\% CL
placed by the neutral detector ND at VEPP-2M using a data sample 
of 2.8~$\pb^{-1}$ \cite{nd1}. Our result supersedes the 
previous value of \mbox{$(0.85 \pm 0.61 \pm 0.12) \cdot 10^{-5}$}
obtained by CMD-2 and based on a data sample 
corresponding to 1.5~$\pb^{-1}$ only \cite{lechner}.
 
The applied procedure of event selection and particularly 
the cut on the angle between the tracks $\Delta{\Psi} < 0.3$
selects events with a rather small $q^2$ so that it is
practically impossible to study the momentum transfer dependence
of the cross section. Therefore, one can conclude 
that the study of transition form factors will become feasible when 
much larger data samples are collected. 

The authors are grateful to the staff of VEPP-2M for the
excellent performance of the collider, to all engineers and 
technicians who participated in the design, commissioning and operation
of CMD-2. We acknowledge useful comments of V.P.~Druzhinin and
stimulating discussions with R.A.~Eichler.

\end{document}